# Seismic Coherent Noise Removal with Residual Network and Synthetic Seismic Simples

Xiao Ma, Gang Yao, Sanyi Yuan, Feng Zhang, Di Wu

*Abstract*—Seismic coherent noise is often found in post-stack seismic data, which contaminates the resolution and integrity of seismic images. It is difficult to remove the coherent noise since the features of coherent noise, e.g., frequency, is highly related to signals. Recently, deep learning has proven to be uniquely advantageous in image denoise problems. To enhance the quality of the post-stack seismic image, in this letter, we propose a novel deep-residual-learning-based neural network named DR-Unet to efficiently learn the feature of seismic coherent noise. It includes an encoder branch and a decoder branch. Moreover, in order to collect enough training data, we propose a workflow that adds real seismic noise into synthetic seismic data to construct the training data. Experiments show that the proposed method can achieve good denoising results in both synthetic and field seismic data, even better than the traditional method.

## I. INTRODUCTION

Post-stack seismic images provide the essential information for unlocking the interior of the Earth and discovering the underground natural resources. Clean seismic images improve the accuracy of aforementioned tasks. However, post-stack seismic images are commonly contaminated by seismic noise, which is caused by many factors including the incompleteness of observed raw seismic data, environment noise and migration algorithms [1]. Effective noise removal, which preserves the amplitude and phase characteristics of signals, boosts the efficiency and accuracy of extracting advanced attributes, as well as, helps to better

Xiao Ma is with State Key Laboratory of Petroleum Resources and Prospecting, China University of Petroleum (Beijing), Beijing 102249, China, and also with College of Geophysics, China University of Petroleum (Beijing), Beijing 102249, China (e-mail: mx297759805@163.com).

Gang Yao is with State Key Laboratory of Petroleum Resources and Prospecting, China University of Petroleum (Beijing), Beijing 102249, China, and also with Unconventional Petroleum Research Institute, China University of Petroleum (Beijing), Beijing 102249, China (e-mail: yaogang@cup.edu.cn)

Sanyi Yuan is with State Key Laboratory of Petroleum Resources and Prospecting, China University of Petroleum (Beijing), Beijing 102249, China, and also with College of Geophysics, China University of Petroleum (Beijing), Beijing 102249, China (e-mail: yuansy@cup.edu.cn).

Feng Zhang is with State Key Laboratory of Petroleum Resources and Prospecting, China University of Petroleum (Beijing), Beijing 102249, China, and also with College of Geophysics, China University of Petroleum (Beijing), Beijing 102249, China (e-mail: zhangfeng@cup.edu.cn).

Di Wu is with State Key Laboratory of Petroleum Resources and Prospecting, China University of Petroleum (Beijing), Beijing 102249, China, and also with College of Geophysics, China University of Petroleum (Beijing), Beijing 102249, China (e-mail: wudi@cup.edu.cn)

understand the targets, e.g., oil and gas reservoirs [2]-[5]. Basically, there are two types of seismic noise: random noise and coherent noise. It is more difficult to suppress the latter than the former. Here, we mainly focus on removing coherent noise in the post-stack seismic images.

Coherent noise, such as ground roll, multiples, and reverberating refractions, is part of the seismic data that mix spatially with signals [6]. Noise generated during the migration process is another type of coherent noise. Traditionally, F-K filters are normally used for coherent noise removal [7], but F-K filters could produce artifacts that reduce the reliability of seismic attributes. There are other methods for the coherent noise removal including Radom transform [11]-[12], Shearlet transform [13]. The migration noise may also be mitigated via aperture optimization [8]-[10], but it necessitates knowledge of local structural dips, and the accuracy of such dip information is critical to the outcome. However, since the coherent noise shares many features with the signals, it is difficult to develop a mathematic method that only remove the undesired noise while retaining the useful reflector signals.

In recent years, deep learning techniques have achieved many successes in seismic denoise, especially for random noise attenuation [14]-[16]. Unlike random noise, however, coherent noise is very difficult to simulate with mathematic formulas. Thus, the training data are an important factor limiting the effectiveness of deep learning methods. To solve this problem, Wang and Nealon [17] proposed a workflow for generating the training data set. By varying the shot numbers between the training input images and the ground truths, the workflow could create the training input-label pairs. But the ground truths still contain some coherent noise using this workflow, which hinders the performance of the deep learning neural network. To further enhance the quality of the training data, Elena *et al*. [18] used the images of dense data as the training data and the images of sparsely subsampled data as the sample. The network trained by these images has a great denoise ability in the experiments of field seismic data.

Besides the training data issue, the network architecture is the other main factor that influences the performance of deep-learning denoise results. DnCNN [19] is widely used for denoise purposes. The network integrates the residual learning strategy and batch normalization operation to effectively remove the random noise in natural images. Wu *et al*. [20] assumed that coherent noise can be simulated by combining Gaussian and Poisson noise; then they design a DnCNN-based network to remove the migration artifacts in field seismic data simultaneously by learning the feature of the mixed noise. But, DnCNN could consume large amounts of computer resources when the network layers become deeper, so there are still problems in the practical application.

In this letter, we introduce a seismic coherent denoise neural network named DR-Unet to enhance the performance on the coherent noise removal in the post-stack seismic images. In DR-Unet, the encoder branch consists of some residual learning blocks to avoid the vanishing gradients problem, and the decoder branch consists of some convolution layers. Moreover, we generate the training data set by mixing real seismic noise with synthetic seismic images. As the synthetic images are noise free, the network can learn the feature of the seismic noise completely. Experiments on both synthetic and field seismic data show the proposed network and workflow are highly effective.

## II. METHODOLOGY

We regard seismic image denoise as a regression problem, which means that the output of the neural network does not need a sigmoid activation function. In other words, instead of probability, the output is the denoised image. In this section, we first introduce the proposed network architecture and the process of generating data. Then, we describe the loss function and measurements for the seismic image denoise task.

### A. Network architecture

We propose a Unet-based network named DR-Unet to mitigate the coherent noise and random noise in seismic images. The input of DR-Unet is a noisy seismic image and the output is a clean seismic image, both of which are fixed as a size of 256×256. As shown in Fig. 1, same as the original Unet, DR-Unet has four encoders and four decoders. The encoders apply convolution kernels followed by max-pool downsampling to encode the input image into feature representations at multiple levels. The decoders semantically project the discriminative features (lower resolution) learned by the encoder onto the pixel space (higher resolution) to get accurately predicted results. The concatenation operation enhances the spatial resolution of the output denoise image since upsampling is a sparse operation and we need a good prior from earlier stages to better represent the localization.

We use the architecture of ResNet-18 [21] to construct each encoder. This is because that ResNets have solved the famously known problem, "vanishing gradient", which happens easily in networks based on Unet. This is because Unet normally has multiple deep layers, and then the gradients from where the loss function is calculated easily shrink to zero after repeated applications of the chain rule. This leads to no learning being performed in the training phase. With ResNets, the gradients can flow directly through the skip connections backward from later layers to initial layers.

The ResNet-18 consists of several residual blocks. Fig. 1(b) shows the architecture of each residual block, which contains three convolution layers implemented by 1×1, 3×3, 1×1 filters, and residual connection operation. Encoders 1-4 (Fig. 1(a)) in our denoise network consist of three residual blocks from ResNet-18, which downsample the input seismic image to 1/2, 1/4, 1/8, 1/16, successively. Then, the output of each encoder block acts as a skip connection for the corresponding decoders.

For the decoder part, we use two convolution layers (Fig. 1(c)) to construct decoders 1-4 (Fig. 1(a)). The kernel size of all the convolution layers in every decoder is 3×3. Note that in the final layer of our denoise network, instead of the sigmoid function used in classification tasks, we use the identity activation function to output the seismic image.

*B. Training data generation and data augmentation*

Before training our network for the seismic image denoise task, we need many 2-D clean seismic data images as ground truth. However, in practice, clean seismic images are very difficult to acquire. To solve this problem, we synthesize the noise-free seismic images as ground truth, which is an ideal label data set for our denoise network. We then add real seismic noise into the synthesized noise-free seismic images to generate the noisy seismic images. As shown in Fig. 2, the procedure of training data set generation are as follows:

First, generating noise-free seismic images. We generate 6400 synthetic 2D seismic images with a size of 256×256. As proposed by [22], to simulate seismic images with realistic geologic structures, fat layers are first generated by creating a sequence of random values in the range [-1,1] that represent the reflection coefficients. Then, some fold and fault structures are added to bend the flat layers, which makes the synthetic seismic data more realistic. Finally, the reflection coefficient sequences are convolved with a Ricker wavelet with a random peak frequency (10~40 Hz), and sample interval (1~8 ms) for the convolution. Again, this workflow produces noise-free seismic images, which are ideal for training labels.

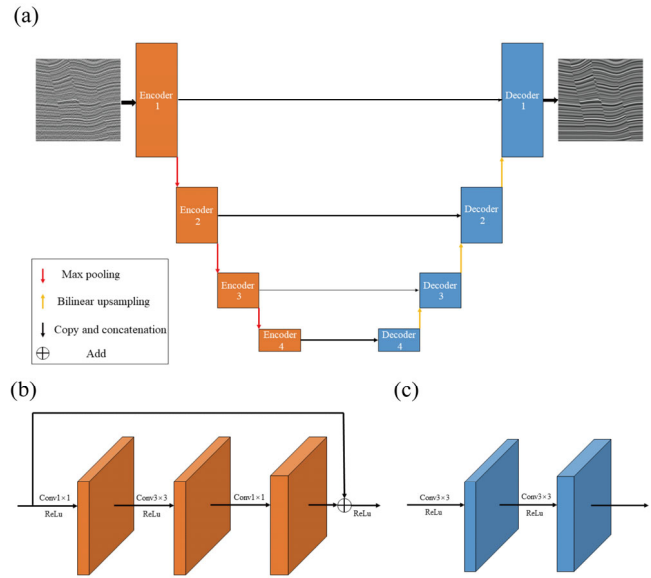

**Fig. 1.** (a) The architecture for DR-Unet consists of an encoder branch and a decoder branch. (b) The architecture of the residual block in each encoder contains three convolution layers and ReLu activation functions. Notes that each encoder contains three

residual blocks in our denoise network. (c) The architecture of each decoder contains two convolution layers and ReLu activation functions.

Second, generating seismic coherent noise samples. The noise data set is extracted from real seismic data (Fig. 2). There are no noticeable real structure events in the noise data set. We divide the noise data set into multiple noise samples with a size of 256× 256. Then weighted noise samples and the clean seismic images are summed up to generate the training dataset. The weighting factor is a scalar $\alpha$ between 0 and 1. Based on our experiments, we set the optimal value of $\alpha$ to 0.55. More details about the weighting factor are discussed in the "Discussion" section.

Third, data augmentation. As shown in Fig. 3, we randomly flip and rotate the existing data samples to increase the number of training data, which can improve the performance of the denoise network.

*C. Loss function and measurement*

We train the network using the *MSE* loss function. We find that the *MSE* loss function could produce stable images. Besides, to evaluate the model accuracy quantitatively, we calculated the peak S/N ratio (*PSNR*) and structural similarity index measure (SSIM) using validation data. The PSNR reflects the amount of noise remaining in the denoised seismic data. *MSE* and *PSNR* can be defined as follows:

$$MSE = \frac{1}{mn}\sum_{i=0}^{m-1}\sum_{j=0}^{n-1}\left|f(i,j) - g(i,j)\right|^2 , \qquad (1)$$

$$PSNR = 10\log_{10}(\frac{MAX_f}{\sqrt{MSE}}) , \qquad (2)$$

where $f(i,j)$ and $g(i,j)$ represent the amplitude of one element of the seismic image with and without noise, respectively, and $MAX_f$ is the maximum amplitude of the clean seismic image. *PSNR* is most used to estimate the effectiveness of the denoise network. The *PSNR* is high when coherent noise is successfully removed.

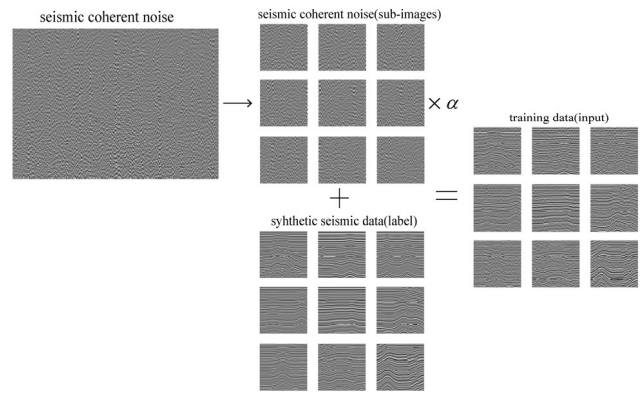

**Fig. 2.** The schematic illustration of the procedure of training dataset generation. The image that only contains seismic coherent noise is divided into sub-images with a size of 256×256; then, the clean synthetic seismic data are mixed with the amplitude-scaled sub-images to produce the training dataset. Note that the clean synthetic seismic data are also used as the ground truth.

*D. Implementation details*

We implement our denoise network using PyTorch, which is a deep-learning library based on Python. The network is trained using an RTX 3090 GPU with 24 GB memory. The weights of ResNet-18 are initialized with the weights after trained on the ImageNet data set. This way enhances the speed of the convergence. The weights of the decoder branch are initialized with random numbers. We update the model parameters using Adam optimizer with a learning rate of 1e-3. The batch size is 20.

During the training process, the loss converges to less than 0.005 after 50 epochs, and the validation loss converges to less than 0.08, which indicates an excellent training process. The SSIM and *PSNR* are stable at 0.94 and 33.5 dB, respectively.

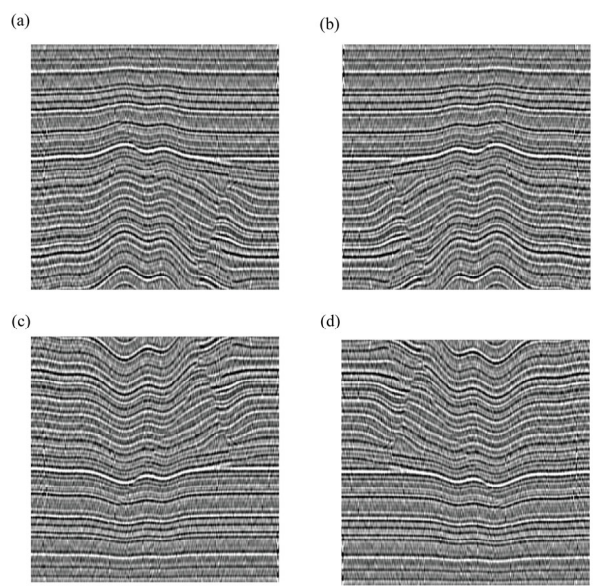

**Fig. 3.** Example of the augmented seismic images. (a) The original seismic image, (b) flipped image (up/down), (c) flipped image (left/right), (d) rotated image.

## III. EXAMPLE

To evaluate the performance of the trained network for coherent noise attenuation, we apply the network to both synthetic and field data sets. In addition, to highlight the effectiveness of the proposed method, we selected the curvelet denoising method for comparison.

### A. Synthetic seismic data test

The synthetic seismic data from the Sigsbee2A model is used to demonstrate the robustness and generalization of our method. This model contains a huge salt, which causes great challenges on seismic migration imaging. Our previous study [23], [24] utilized sparsity-constrained least-squares reverse-time migration to produce a state-of-art image, which shown in Fig. 4(a). This image shows great details of the subsalt section. However, one prominent flaw is that the coherent noise, i.e., migration artifacts, is prevail, some of which are indicated by the arrows. Unfortunately, it is almost impossible to eliminate this type of noise with conventional denoise filters, e.g., curvelet denoising (Fig. 4(b)), because noise shares similar frequency and wavenumber features with the signals, i.e., real geologic reflectors.

As shown in Fig. 4(c), the trained network can suppress the migration artifacts while preserving the integrity of the geological reflectors. The reflectors become more continuous, and no longer are contaminated by the migration artifacts. More specifically, as indicated by the black arrows, the trained network can attenuate the migration artifacts in sediment and salt body. However, the curvelet method (Fig. 4(b)) fails to remove the noise in these regions. The result implies that the learned features could be used to distinguish geologic structures from the migration artifacts. One explanation can be speculated from Fig. 4(e): the migration artifacts have different shape patterns from the geologic reflectors, i.e., the migration artifacts have the arc-shape appearance while the geologic reflectors are flatter and more continuous. These pattern features cannot be distinguished by a conventional filter like curvelet denoising based on frequency and wavenumber characters.

### B. Field seismic data test

Although the network is trained using only synthetic data sets, it also works for field data. In this section, we test the trained network on a typical real seismic image to further demonstrate its powerful performance. The real seismic images are normalized in amplitude before fed to the network, but the amplitude is scaled back after denoise.

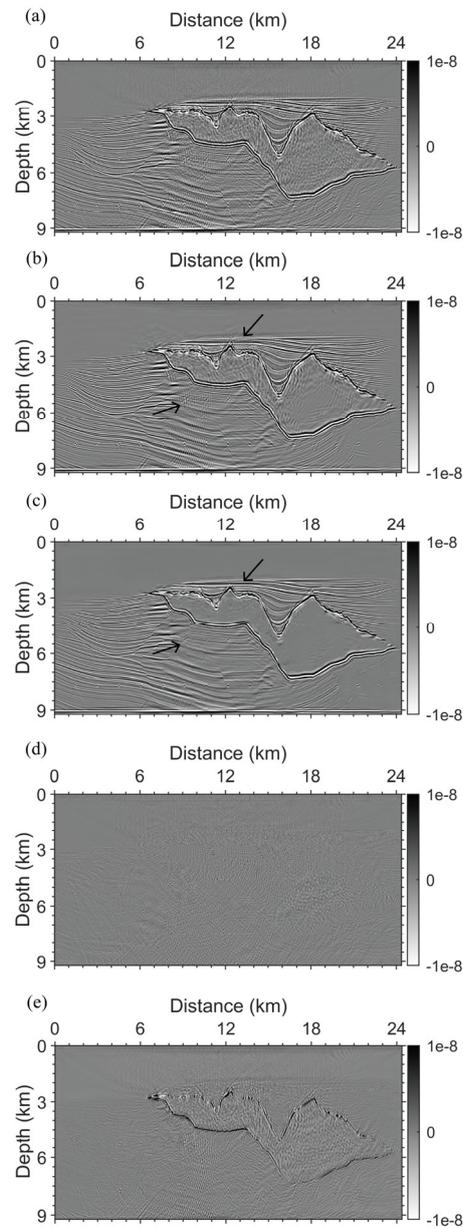

**Fig. 4.** (a) The migration image of the Sigsbee2A model. (b) Denoise result with curvelet denoising. (c) Denoise result with the trained network. The arrows highlight the noise removed by the trained network but not the curvelet denoising. (d) The removed noise computed by subtracting (b) from (a). (e) The removed noise computed by subtracting (c) from (a).

As shown in Fig. 5(a), the post-stack seismic image contains strong and omnipresent coherent noise. The noise distorts the seismic reflectors, which may cause mis-interpretation, resulting in drilling risks. We apply the trained DR-Unet to attenuate seismic coherent noise. As shown in Fig. 5(c), it is obvious that the network has successfully removed the coherent noise. Fig. 5(e) shows the removed noise by the network. As can be seen, there is no geological reflectors removed from the seismic image, which demonstrates a good amplitude-preserving ability. However, as shown in Fig. 5(b), curvelet denoising blurs geological features such as faults, which hinders the process of interpretation. Besides, the trained DR-Unet removes the random noise in the real seismic images. We also applied the trained network on many other real data sets. All experiments imply that the trained DR-Unet performs well on both coherent and random noise removal in both synthetic and field data sets.

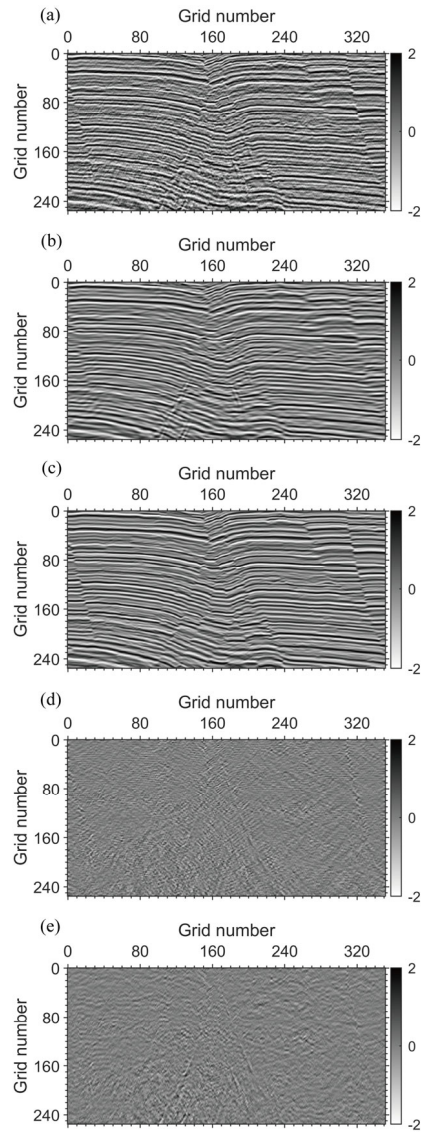

**Fig. 5.** (a) The field seismic image that contains coherent noise. (b) Denoise result with curvelet denoising. (c) Denoise result with the trained network. (d) The removed noise computed by subtracting (b) from (a). (e) The removed noise computed by subtracting (c) from (a).

## IV. Discussion

The performance of deep-learning networks highly depends on training data. During this study, we found that the weighting factor $\alpha$ affects the performance of the DR-unet. The smaller this factor, the weaker the noise in the sample, and vice versa. As shown in Fig. 6(b) and (c), when the value of $\alpha$ is 0.2, we find that the denoise network has a good ability to preserve the amplitude of the reflectors. However, the network fails to mitigate most of the coherent noise in the image. By contrast, as shown in Fig. 6(d) and (e), the network attenuates almost all the coherent noise but also suppresses some reflections when the value of $\alpha$ is 1. Based on our tests, we found that the optimal value of $\alpha$ is 0.55 for our denoise test.

We conclude that a high percentage of noise in training data can cause neural network to classify the reflections as part of the coherent noise; on the contrary, a low percentage of noise may be not enough for the network to learn the feature of the seismic coherent noise.

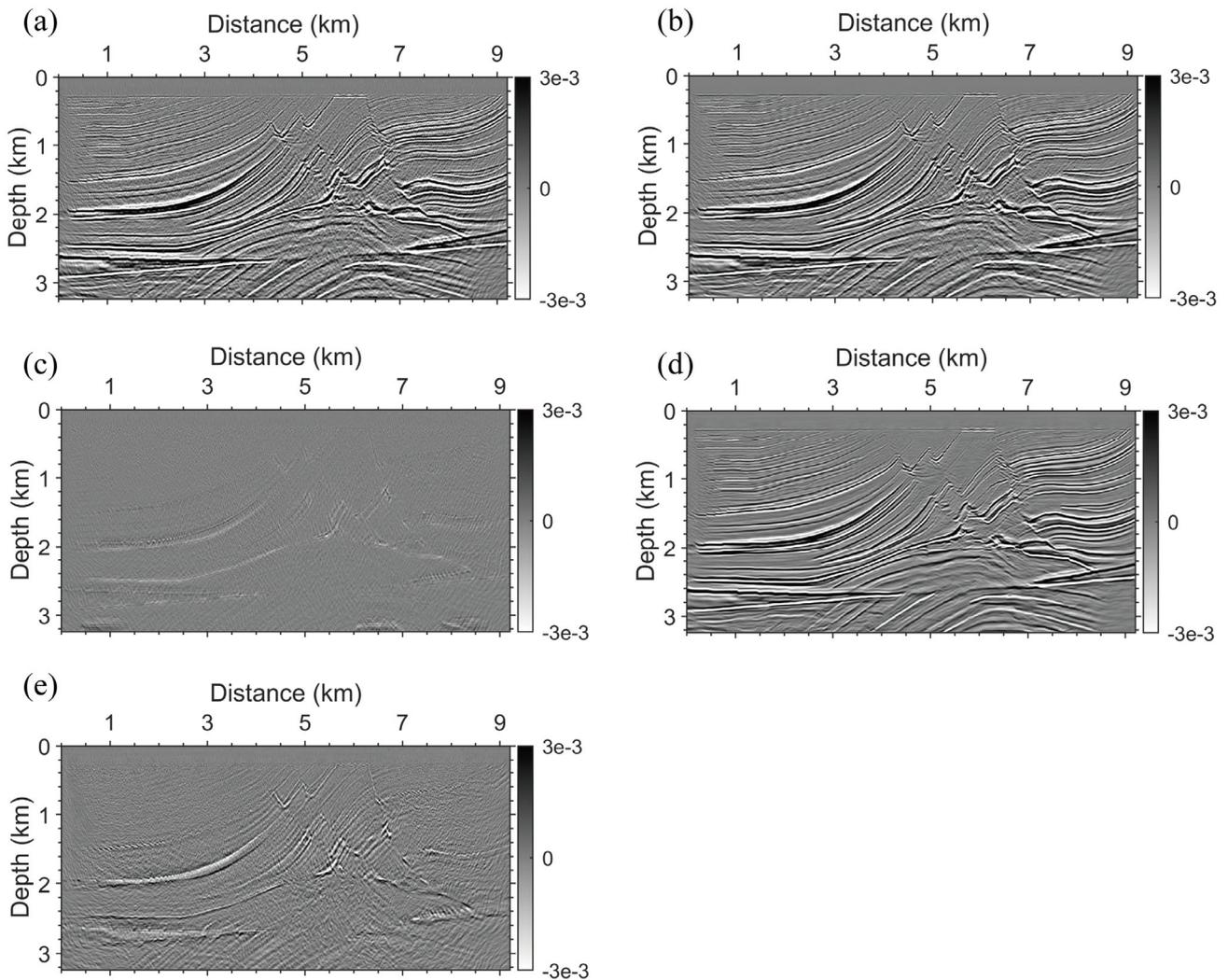

**Fig. 6.** The influence of the weight of noise in training data. (a) the migration image of the Marmousi model that contains coherent noise. (b) The denoised result of our neural network when $\alpha = 0.2$. (c) The removed noise when $\alpha = 0.2$. (d) The denoised result of our neural network when $\alpha = 1$. (e) The removed noise when $\alpha = 1$.

## V. CONCLUSION

In this letter, we propose a novel CNN-based network named DR-Unet to attenuate the coherent noise in synthetic and field post-stack seismic data. DR-Unet consists of an encoder branch and a decoder branch. The encoder branch is designed based on Resnet18 to avoid the gradient vanish problem. To mitigate the impact of inadequate training data set, we combine the real seismic noise with synthetic seismic data to form the training data set. Many synthetic and field data tests demonstrate that our method removes the coherent noise effectively while preserving the reflectors.

## REFERENCES

[1] G. H. Gardner and A. Canning, "Effects of irregular sampling on 3-D prestack migration," in *Proc. SEG Tech. Program Expanded Abstr. Soc. Explor. Geophysicists*, 1994, pp. 1553-1556.

[2] K. J. Marfurt, R. L. Kirlin, S. L. Farmer, and M. S. Bahorich, "3-D seismic attributes using a semblance-based coherency algorithm," *Geophysics*, vol. 63, no. 4, pp. 79-88, Jul. 1998.

[3] Q. Gan, and D. Elsworth, "Analysis of fluid injection-induced fault reactivation and seismic slip in geothermal reservoirs," *Journal of Geophysical Research: Solid Earth*, vol. 119, no. 4, pp. 3340-3353, Mar. 2014.

[4] S. Y. Yuan, J. Liu, S. Wang, T. Wang, and P. Shi, "Seismic waveform classification and first-break picking using convolution neural networks," *IEEE Trans. Geosci. Remote Sens.*, vol. 15, no. 2, pp. 272-276, Jan. 2018.

[5] Q. He, Y. Wang, "Reparameterized full-waveform inversion using deep neural networks," *Geophysics*, vol. 86, no. 1, pp. 1-13, Jan. 2021.

[6] K. Verma, S. Guo, T. Ha, and K. J. Marfurt, "Highly aliased ground-roll suppression using a 3D multiwindow Karhunen-Loeve filter: Application to a legacy Mississippi Lime survey," *Geophysics*, vol. 81, no. 1, pp. 79-88, Jan. 2016.

[7] D. Y. Wang and Y. Ling, "Phase-shift-and phase-filtering-based surface-wave suppression method," *Applied Geophysics*, vol. 13, no. 4, pp. 614-620, Jan. 2017.

[8] S. Xu, Y. Zhang, and B. Tang, "3D angle gathers from reverse time migration," *Geophysics*, vol. 76, no. 2, pp. 77-92, Mar. 2011.

[9] A. Klokov and S. Fomel, "Selecting an optimal aperture in Kirchhoff migration using dip-angle images," *Geophysics*, vol. 78, no. 6, pp. 242-254, Oct. 2013.

[10] M. Yu, X. Gong, and X. Wan, "Seismic Coherent Noise Removal of Source Array in the NSST Domain," *Appl. Sci.*, vol. 12, no. 21, Oct. 2022.

[11] Y. Luo, J. Xia, R. D. Miller, Y. Xu, J. Liu, and Q. Liu, "Rayleigh-wave dispersive energy imaging using a high-resolution linear Radon transform," *Pure Appl. Geophy.*, vol. 165, no. 5, pp. 903-922, May. 2008.

[12] B. Abbad, B. Ursin, and M. J. Porsani, "A fast, modified parabolic Radon transform," *Geophysics*, vol. 76, no. 1, pp. 11-24, Jan. 2011.

[13] M. Yu, X. Gong, and X. Wan, "Seismic Coherent Noise Removal of Source Array in the NSST Domain," *Applied Sciences*, vol. 12, no. 21, Oct. 2022

[14] S. Yu, J. Ma, and W. Wang, "Deep learning for denoise," *Geophysics*, vol. 84, no. 6, pp. 333-350, Nov. 2019.

[15] N. Liu, J. Wang, J. Gao, S. Chang, and Y.Lou, "Similarity-Informed Self-Learning and Its Application on Seismic Image Denoising," *IEEE Trans. Geosci. Remote Sens.*, vol. 60, Sep. 2022.

[16] J. Li, X. Wu, and Z. Hu, "Deep learning for simultaneous seismic image super-resolution and denoising," *IEEE Trans. Geosci. Remote Sens.*, vol. 60, pp. 1-11, Feb. 2021.

[17] E. Wang, and J. Nealon, "Applying machine learning to 3D seismic image denoising and enhancement," *Interpretation*, vol. 7, no. 3, pp. 131-139, Aug. 2019.

[18] K. Elena *et al.*, "Leveraging deep learning for seismic image denoising," *First Break*, vol. 38, no. 7, pp. 41-48, Jul. 2020.

[19] K. Zhang *et al.*, "Beyond a gaussian denoiser: Residual learning of deep cnn for image denoising," *IEEE Trans. Image Process.*, vol. 26, no. 7, pp. 3142-3155, Feb. 2017.

[20] H. Wu, B. Zhang, and N. Liu, "Self-adaptive denoising net: Self-supervised learning for seismic migration artifacts and random noise attenuation," *J. Pet. Sci. Eng.*, vol. 214, Jul. 2022.

[21] K. He, X. Zhang, S. Ren, and J. Sun, "Deep residual learning for image recognition," in *CVPR*, pp. 770-778, 2016.

[22] X. Wu, L. Liang, Y. Shi, and S. Fomel, "FaultSeg3D: Using synthetic data sets to train an end-to-end convolutional neural network for 3D seismic fault segmentation," *Geophysics*, vol. 84, no. 3, pp. IM35-IM45, May. 2019.

[23] Wu B., Yao G., Cao J. J., et al., "Huber inversion-based reverse-time migration with de-primary imaging condition and curvelet-domain sparse constraint," *Pet. Sci.*, vol.19, no. 4, pp. 1542-1554, Aug. 2022.

[24] G. Yao, B. Wu, N. V. da Silva, H. A. Debens, D. Wu, and J. Cao, "Least-squares reverse-time migration with a multiplicative Cauchy constraint,". *Geophysics*, vol. 87, no. 3, pp. 1MJ-V246, Jun. 2022.